\newcommand{\Msun}{~M_\odot}

\newcommand{\kms}{\rm ~km~s^{-1}}

\newcommand{\ml}{~\Msun ~\rm yr^{-1}}

\documentclass[12pt,preprint]{aastex}

\begin{document}

\title{CASSIOPEIA A AND ITS CLUMPY PRESUPERNOVA WIND}
\author{Roger A. Chevalier and Jeffrey Oishi}
\affil{Department of Astronomy, University of Virginia, P.O. Box 3818, \\
Charlottesville, VA 22903; rac5x@virginia.edu, jo8c@virginia.edu}


\begin{abstract}
The observed shock wave positions and expansion in Cas A
can be interpreted in a model of supernova interaction with a freely
expanding stellar wind with a mass loss rate of $\sim 3\times 10^{-5}\ml$
for a wind velocity of $10\kms$.
The wind  was probably still being lost at the time of the supernova,
which may have been of Type IIn or IIb.  
The wind   
may play a role in the formation of very fast knots observed in Cas A.
In this model, the quasi-stationary flocculi (QSFs) represent clumps in the
wind, with a density contrast of several $10^3$ compared to the smooth wind.
The outer, unshocked 
clumpy wind is photoionized by radiation from the supernova,
and is observed as a patchy HII region around Cas A.
This gas has a lower density than the QSFs and is heated
by nonradiative shocks driven by the blast wave.
Denser clumps have recombined and are observed as HI compact absorption
features towards Cas A.

\end{abstract}


\section{INTRODUCTION}

The supernova remnant Cas A (Cassiopeia A) gives us our best
view of the outcome of the explosion of a massive star.
Spectral imaging with {\it Chandra} at X-ray wavelengths 
(Hughes et al. 2000) and
{\it HST} at optical wavelengths (Fesen et al. 2001) has
shown the complex structure of the ejected heavy elements.
The {\it Chandra} image also revealed a central compact
X-ray source (Tananbaum 1999), probably a neutron star,
and lines of the radioactive isotope $^{44}$Ti have been
detected (Iyudin et al. 1994).
Despite these many developments, the evolutionary status of
Cas A  remains uncertain.
The most common assumption is that the supernova is interacting
with a constant density interstellar medium (Gull 1973b; Gotthelf
et al. 2001; DeLaney \& Rudnick 2003), or perhaps with a molecular 
cloud (Keohane, Rudnick, \& Anderson 1996).
Interaction with a circumstellar shell has also been suggested
(Chevalier \& Liang 1989; Borkowski et al. 1996).
The immediate environment of a massive star is expected to be
strongly influenced by mass loss, and the pervasive, high-velocity, heavy element
ejecta in Cas A indicate that the star underwent strong mass loss
before the explosion.

Here, we propose that the supernova is interacting with the
slow wind from the progenitor star, with a $\rho_w\propto r^{-2}$
density profile.
The resulting model can be compared to the width and expansion
of the shocked region (\S~2), giving constraints on the basic parameters.
Implications of the model for inhomogeneities in the 
the wind, for the supernova, and for  a surrounding HII region
and HI knots are discussed in \S~3.

\section{WIND INTERACTION MODEL}

The distance to Cas A has been determined from the fast knot expansion
to be $3.4^{+0.3}_{-0.1}$ kpc (Reed et al. 1995).
Ashworth (1980) claimed an observation of the Cas A supernova by
Flamsteed in 1680, but that claim has been controversial (Stephenson \& Green 2002).
On the basis of very fast knots that show little sign of deceleration,
Thorstensen et al. (2001) determined an explosion date of $1671.3\pm 0.9$.
We take an explosion date of $1675\pm 5$.
The outer shock front has been clearly observed in {\it Chandra} images to
have a radius of 153$^{\prime\prime}$ (Gotthelf et al. 2001),
or $7.8\times 10^{18}$ cm at a distance of 3.4 kpc.
The position of the reverse shock front is less clear, but was determined
by Gotthelf et al. (2001) from the  inner edge to the bright ring
of emission at X-ray and radio wavelengths; they found a
ratio of the forward shock radius to that of
the reverse shock of $r_f/r_r=1.5$ with a variation of 14\% around the remnant.

The youth of Cas A has enabled studies of its expansion from
proper motion studies.
DeLaney \& Rudnick (2003) have recently measured the expansion
of the forward shock in X-rays from {\it Chandra} observations over
2000--2002 and found it to be in the large range $0.02-0.33$ \%~yr$^{-1}$,
with a median of $0.21$ \%~yr$^{-1}$.
The median corresponds to an expansion parameter $m_f=d\ln r_f/d\ln t$ of 
$0.68$.  
The bright ring of X-ray emission has previously been found to
be expanding at $0.20\pm 0.01$ \%~yr$^{-1}$ from {\it Einstein} and
{\it ROSAT} observations covering 1979--1996 (Koralesky et al. 1998; Vink et al. 1998),
or $m=0.62\pm 0.03$.
The bright radio ring is approximately co-extensive with the X-ray one.
Ag\"ueros \& Green (1999) studied the minima in the visibility plane at 151 MHz
over the period 1984--1997 to determine a timescale for the bulk ring expansion
of $460\pm 30$ years, or $m=0.69\pm 0.05$.
While this result is consistent with the X-ray expansion, other radio studies
have yielded a slower expansion;  Anderson \& Rudnick (1995)
find an expansion age of $750-1300$ years ($m=0.33\pm 0.11$).
DeLaney \& Rudnick (2003) recently examined the motion of the radio
ring with an emphasis on angle-averaged emissivity profiles and
found an expansion of $0.07\pm 0.03$  \%~yr$^{-1}$
($m=0.22\pm 0.09$).
DeLaney \& Rudnick suggested the difference is due to the more rapid flux drop
of the ring compared to the plateau, but this is not definitively established.
We regard the current situation on the radio ring expansion to be uncertain;
more observations are needed.

We have carried out simulations of supernova interaction to compare
to these observations of the shock dynamics.
The explosion of Cas A appears to have been that of a massive star
core, even though there was some H near the surface (Fesen \& Becker 1991).
We thus use the model of Matzner \& McKee (1999) for the density
distribution resulting from the explosion of a massive star with
a radiative envelope.
We concentrate on interaction with a stellar wind from the progenitor
star with density $\rho_w=Ar^{-2}$, although we also briefly
consider a constant density environment.
For a steady wind, $A=\dot M/4\pi v_w$, where $\dot M$ is the mass
loss rate and $v_w$ is the wind velocity.
The outer part of the Matzner \& McKee (1999) profile has the
form $\rho_{sn}\propto r^{-10.12}$.
The self-similar solutions of Chevalier (1982) show that when such
a profile interacts with a wind, both reverse and forward shocks
expand with $m=0.88$ and the thickness of the shocked region is
$r_f/r_r=1.26$.
The observed shock parameters indicate that the reverse
shock wave has propagated
in from the power law region.

In order to calculate the further evolution of the shock fronts,
we used the VH-1 hydrodynamics code to compute the 1-dimensional
evolution of the wind interaction flow.
In order to take advantage of the scaling that applies to this
problem (Gull 1973a; Truelove \& McKee 1999), we used the
dimensionless variables $r'=r/R'$, $v'=v/V'$, and $t'=t/T'$,
where $R'=M_{ej}/(4\pi A)$, $V'=(2E/M_{ej})^{1/2}$,  $T'=R'/V'$,
$M_{ej}$ is the ejecta mass, and $E$ is the explosion energy.
We have $R'=3.16\times 10^{19}M_1 A_{-5}^{-1}$ cm,
$V'=3160 E_{51}^{1/2}M_1^{-1/2}\kms$, and
$T'=3160 E_{51}^{-1/2}M_1^{3/2}A_{-5}^{-1}$ yr,
where $M_1$ is the ejecta mass in units of $10\Msun$,
$E_{51}$ is the energy in units of $10^{51}$ ergs,
and $A_{-5}$ is $A$ in units of $10^{-5}\ml/(4\pi 10\kms)$.
Fig. 1 shows  
that $r_f/r_r=1.5$ when $t'=1.56$ in the scaled variables.
At this time, the forward shock has an expansion parameter $m_f=0.76$
and the reverse shock has $m_r=0.68$; the forward shock radius is $r'_f=1.48$.
In a computation with constant density ejecta, we found similar values
of the $m$ parameters when $r_f/r_r=1.5$; the sensitivity to the
supernova density profile is weak.
Using the Matzner \& McKee (1999) supernova density profile,
we have also carried out a  computation for expansion in
a uniform (interstellar) medium.
In this case, when $r_f/r_r=1.5$, the forward shock 
has an expansion parameter $m_f=0.50$
and the reverse shock has $m_r=0.27$.

For the forward shock motion, the wind model gives an expansion rate
of 0.235 \% yr$^{-1}$, while the uniform model gives 0.153 \% yr$^{-1}$.
The wind value appears to better represent the observations, which
have a median value of 0.21 \% yr$^{-1}$ (see Fig. 3 of
DeLaney \& Rudnick 2003).
As discussed above, there is  ambiguity in the motion of the bright
ring, which is identified as gas that has passed through 
and is bounded by the reverse shock.
The X-ray data and some radio data are consistent with the wind interaction
model, while other radio data (expansion of 0.07 \% yr$^{-1}$ found
by DeLaney \& Rudnick 2003) are consistent with a constant density
surroundings.
In the wind model,
the small rate of expansion of compact radio features and parts of
the forward shock front may be due to interactions with dense
inhomogeneities in the wind like the QSFs (quasi-stationary 
flocculi; Van den Bergh 1971b).
Despite the current small rate of expansion for some segments
of the forward shock, the shock is fairly circular overall
(Gotthelf et al. 2001), suggesting that the slowing is due to
recent interaction with clumps.

The radial emissivity profiles are another difference between
the models.
Compared to the uniform density case, the shocked ejecta are concentrated
into a higher density region surrounded by a region of near constant
density wind in the wind case (Fig. 2).
The ejecta shell in the wind case is broadened by a factor $\sim 3$
by hydrodynamic instabilities (see Fig. 8 of Chevalier, Blondin, \& Emmering
1992), which will be crucial for interpreting the spatial distribution
of X-ray emission from Cas A.
Radial profiles of radio and Si emission (Fig. 4 of Gotthelf et al. 2001)
 show an outer plateau of emission, but more detailed
investigations of the observations, together with multidimensional
hydrodynamic models are needed to provide firm results.

Adopting the wind model, we can apply two known properties of
Cas A, its age $t=320$ yr and outer shock radius $r_f=7.8\times 10^{18}$ cm, to
determine relations between the 3 model parameters.
We find $M_1=0.16\Msun E_{51}$ and $A_{-5}=1.3E_{51}$.  
The uncertainties in the shock positions give an uncertainty in the
numerical coefficients of $\sim 50$\% and additional physical effects
that could affect the hydrodynamics, such as cosmic ray pressure or
clumpiness, increase the uncertainty.
As an example, we take $E_{51}=2$, leading to an ejecta mass of $3.4\Msun$
(mostly heavy element core material)
and a mass loss rate of $2.6\times 10^{-5}\ml$ for a wind velocity of $10\kms$.
The current mass of shocked wind material is $6.4\Msun$ and the shocked
ejecta mass is 1.9 $\Msun$.
Including a neutron star mass of $1.4\Msun$ brings the core mass to
$4.8\Msun$, 
corresponding to a main sequence mass of $\sim 17\Msun$.

Another constraint on the models comes from the X-ray luminosity,
which is related to the X-ray emitting mass.
Mass estimates include $\ga 15\Msun$ from {\it Einstein} data (Fabian et al. 1980),
$\sim 14\Msun$ 
from {\it ASCA} data
(Vink, Kaastra, \& Bleeker 1996), and $10\Msun$ from {\it XMM-Newton}
data (Willingale et al. 2002).
The model described above is approximately consistent with these results.
The model cannot be expected
to yield improved values for the ejecta mass and explosion energy.
The main point is that the dynamics and emission suggest that the
supernova is interacting with a moderately dense stellar wind, considerably
denser than the wind expected from a Wolf-Rayet star,
which typically have $\dot M\approx 10^{-5}\ml$ and
$v_w\approx 10^3\kms$, but consistent
with the wind from a red supergiant star.

\section{IMPLICATIONS}

\subsection{Wind Inhomogeneities}

The presence of QSFs 
was one of the reasons that Chevalier \& Liang (1989) used to argue
for a dense circumstellar shell.
The optical emission from QSFs indicates that it is from 
radiative shock fronts
with velocities $v_q\sim 100-200\kms$.
The velocity of the forward shock front is $v_f=5800\kms$, so the
preshock density in the QSFs is $n_q\approx n_0(v_f/v_q)^{2}
=3\times 10^3 n_0$, where $n_0$ is the smooth wind density;
in our model, the density at the shock front is 
currently $\sim 1$ H atom cm$^{-3}$.
The high density contrast is suggestive of a shell, but the positions
of the QSFs are not restricted to the bright emitting shell of Cas A
(see Fig. 3 of Lawrence et al. 1995 and Fig. 10 of Fesen 2001).
This implies that the QSFs are dense clumps within a smoother
wind with the properties given above.
The presence of wind inhomogeneities is also indicated by the
irregular outline of the forward shock front observed
in X-rays (Gotthelf et al. 2001).  

The maximum shock velocity in the QSFs may be determined by the cooling time
in the postshock region.
A similar situation may be present in the remnant of SN 1987A, and
Pun et al. (2002) estimate the postshock cooling time as
$t_{\rm cool}\approx 2.2 (2\times 10^4{\rm~amu~cm^{-3}}/\rho_q) 
\linebreak[0] (v_q/250 \kms)^{3.8}$ yr over the 
shock velocity range $100-600\kms$.
Converting from $\rho_q$ to the ambient preshock density, $\rho_0$, 
the ram pressure relation above gives $t_{\rm cool}\propto v_q^{5.8}$.
Substituting the current conditions for the blast wave
yields $v_q =280 (t_{\rm cool}/100 {\rm~yr})^{0.17}\kms$.
For higher velocities, the shocks are nonradiative, which explains why
this is approximately the upper limit of the shock wave velocities in
the QSFs.
There may be faster shock waves moving into lower density inhomogeneities,
which are not visible optically.

The origin of high contrast knots in the circumstellar wind is not clear,
but they have probably been observed on other objects.
In SNe IIn (Type IIn supernovae), 
there is moderately narrow line emission, probably from shocked
clumps, as well as broad line emission;  for example, SN 1988Z showed
broad H$\alpha$ with velocities to $20,000\kms$ as well as a narrower
$2000\kms$ component (Stathakis \& Sadler 1991).
Radiative shocks are present at higher velocities than in
Cas A because of the higher ram
pressure at early times in the supernova evolution.
SN 1995N is another SN IIn with an intermediate width
H$\alpha$ component as well as narrow lines that appear to be
from clumps in the preshock circumstellar medium (Fransson et al. 2002).

\subsection{The Supernova}

The presupernova star apparently had  little H at the time of the explosion,
but did have some (Fesen \& Becker 1991; Fesen 2001).
This implies that the supernova may have been more closely related
to Type IIn and IIb supernovae than to Type Ib and Ic supernovae,
which are probably the explosion of Wolf-Rayet stars, and that
the dense circumstellar wind around Cas A may have
extended down to close to the stellar surface.
The Type IIn  SN 1995N showed evidence for interaction
with a dense, H-rich wind, but also for emission
from fast, O-rich ejecta near the reverse shock, showing that
the explosion occurred with little H at the surface of the star
(Fransson et al. 2002).

The Type IIb SN 1993J also had little H at the time of the explosion
and expanded into a dense wind.
Radio and X-ray observations imply a mass loss rate of $\sim 4\times 10^{-5}\ml$
for a wind velocity of $10\kms$ (Fransson, Lundqvist, \& Chevalier 1996),
comparable to our estimate for Cas A.
In addition, the main sequence mass of the progenitor star is estimated
at $13-16\Msun$ (Woosley et al. 1994), close to our estimate for Cas A.
Houck \& Fransson (1996) estimated that the bulk of the H/He envelope
mass of SN 1993J lies between $8500-10,000\kms$;
although most of the knots in Cas A do not show H lines,
Fesen \& Becker (1991) found a H knot with a velocity $\sim 9000\kms$.
He enrichment is found in both the QSFs (Chevalier \& Kirshner 1978) and
in the H envelope of SN 1993J (Houck \& Fransson 1996).
A possible problem is that some models for SN 1993J require a massive
star companion, which survives the supernova (Woosley et al. 1994);
there is no evidence for a massive star near the explosion center of Cas A
(Thorstensen et al. 2001).
However, it may be possible for a star to undergo this evolution as
a single star.

If the dense wind initially extended in to close to the stellar surface,
it can help to explain a puzzling feature of the fastest knots.
Fesen \& Becker (1991) find N and H rich knots moving at
$\sim 10,000\kms$, which is surprising because the outer supernova
ejecta are expected to be shocked to a high temperature and have
a long radiative lifetime.
These knots must have crossed the reverse shock  early in
the life of the supernova remnant.
In the dense, slow wind, the early evolution is given by a
self-similar solution, with $r_f/t=30,000E_{51}^{0.44}M_1^{-0.32}
A_{-5}^{-0.12}(t/{\rm day})^{-0.12}\kms$.  
For the typical parameters, ejecta moving at $10,000\kms$ 
crosses the reverse shock at
an age of $\sim 25$ years, when the preshock density is $2\times 10^{-21}$
g cm$^{-3}$.
The smooth, H-rich ejecta are not radiative at this time, but a moderate degree
of clumping can lead to radiative cooling and knot formation.
The absence of a dense wind would lead to hotter, lower density ejecta
and would make it difficult to produce
cool ejecta knots at this early time.

\subsection{The Outer Wind}

A early photograph of Cas A taken by Minkowski, reproduced
in van den Bergh (1971a), gives evidence for a patchy HII region
surrounding Cas A, extending out 7$^{\prime}$ from the remnant
(see also Fig. 10 of Fesen 2001).
Spectral observations of the nebulosity on the E side of Cas A by
Fesen, Becker, \& Blair (1987) showed it to be a low-ionization
HII region or shock-heated gas; we advocate the HII region interpretation
here.
Peimbert \& van den Bergh (1971) estimated an intrinsic emission
measure $\sim 2500$ pc cm$^{-6}$, 
leading to an electron density $n_e\approx 15$ cm$^{-3}$
for a radius of 5.7 pc.
The corresponding mass is several $100\Msun$, which is too high
for the material to be stellar mass loss.
However, the emission is patchy, and we suggest that the emitting
gas is in clumps with $n_e\sim 300$ cm$^{-3}$, reducing the mass by
a factor $\sim 20$.
The recombination time for the gas is $\sim 300$ years, so it is
possible that it was ionized at the time of the supernova shock
breakout; the current X-ray luminosity of Cas A is not sufficient
to provide the ionization.
As with the photoionized gas around SN 1987A (Lundqvist \& Fransson 1996),
the dominant density component that is observed is the densest one
that has not already cooled and recombined.
Peimbert \& van den Bergh (1971) estimated that an energy in ionizing
radiation of $1\times 10^{50}$ ergs was needed to ionize the gas.
This is larger than expected during shock breakout.
With our reduced mass, the energy requirement drops to 
$\sim 5\times 10^{48}$ ergs,
which can be attained at shock breakout for an extended star with
a low mass envelope (Matzner \& McKee 1999).

In this scenario, the patchy HII region represents mass lost
from the progenitor during a red supergiant phase.
At $10\kms$, the wind can reach 6.9 pc (7$^{\prime}$ at 3.4 kpc) in
$7\times 10^5$ yr.
This is approximately the expected age of the red supergiant phase
for a $15-20\Msun$ mass star.
The extended red supergiant wind is apparently also observed around
SN 1987A, although by a dust echo in this case (Chevalier \& Emmering 1989).
The wind ends in a patchy shell of radius 4.5 pc.
A termination shell might also be present around Cas A.

Knots in the preshock wind may have also been detected as HI
compact absorption features towards Cas A (Reynoso et al. 1997);
the knots have sizes $<0.1$ pc and show spatial substructure.
Their densities are presumably $\ga 300$ cm$^{-3}$ so
that the H can recombine.
The knots have radial velocities in the range $-10$ to $-17\kms$
relative to the systemic velocity of Cas A (Reynoso et al. 1997).
In the present model, 
these velocities represent the velocity of the presupernova wind
and are consistent with the wind of a red supergiant star.

In summary, we have shown that the dynamical properties of Cas A are
consistent with interaction with the dense wind from a red supergiant
progenitor star.
The wind interaction supports a Type IIb or IIn supernova designation,
which is also indicated by the presence of high velocity hydrogen
(Fesen \& Becker 1991).
The basic emission features of the supernova remnant, fast ejecta (fast moving
knots), intermediate velocity shocks (quasi-stationary flocculi), and
narrow line emission (surrounding ionized clumps), have also been
identified as emission features in Type IIn supernovae.
The ability to spatially resolve the complex interaction in Cas A
may provide a useful guide to the interpretation of the distant
supernova emission.
Finally, we note that some of the points of view advocated here, including
interaction with a red supergiant wind and the possible importance
of wind interaction for the formation of fast knots, have recently
been discussed by Laming \& Hwang (2003).

\acknowledgments
We are grateful to John Blondin for assistance in using the VH-1 code, to
Rob Fesen for comments on the manuscript,
and especially to Larry Rudnick
for a helpful referee's report.
This research was supported in part by   NSF grant AST-0307366.

\clearpage

\begin{figure}[!hbtp]
\plotone{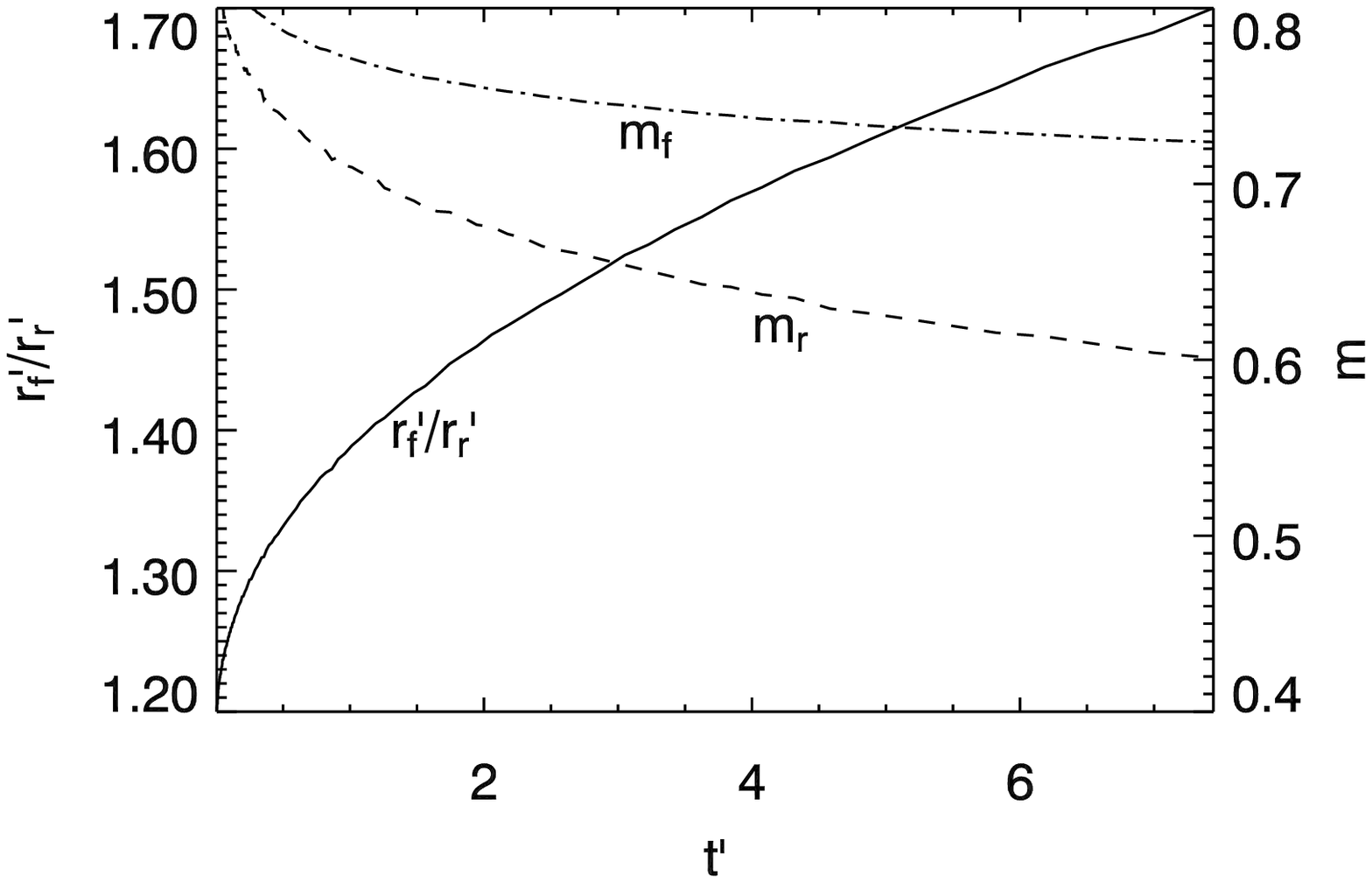}
\figcaption{The ratio of forward shock to reverse shock radius,
$r'_f/r'_r$, and the deceleration parameters for the forward
shock, $m_f$, and reverse shock, $m_r$, as a function of scaled
time.  The hydrodynamic model is for a exploded star with a radiative
envelope running into a wind with a $\rho_w\propto r^{-2}$ density profile.}
\end{figure}

\clearpage

\begin{figure}[!hbtp]
\plotone{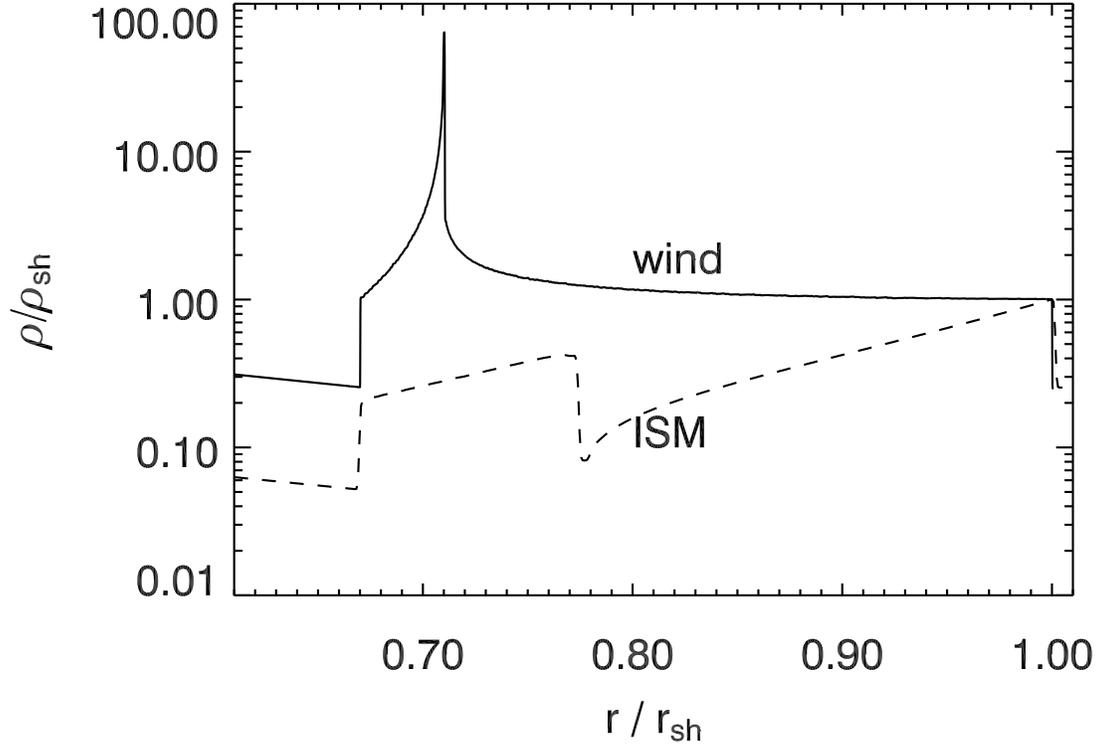}
\figcaption{The density profile labeled `wind' is the same model
as in Fig. 1 shown when $r'_f/r'_r=1.5$.
The `ISM' model has the same supernova model, but is running into a
constant density medium.
The density and radius are scaled to the values at the outer shock front.
The value of $r'_f/r'_r$ is chosen to be close to that observed in Cas A.}
\end{figure}

\end{document}